\documentclass[aps,prl,twocolumn,superscriptaddress,numerical]{revtex4-2}

\usepackage[colorlinks,
citecolor=blue,
linkcolor=blue,
urlcolor=magenta]{hyperref}

\usepackage[utf8]{inputenc}
\usepackage[english]{babel}
\usepackage[T1]{fontenc}
\usepackage{amsmath,amssymb,dsfont,orcidlink}
\usepackage[normalem]{ulem}

\DeclareMathAlphabet{\mathcal}{OMS}{cmsy}{m}{n}

\newcommand{\prlsection}[1]{\textbf{\textit{#1.}}}

\newcommand{\E}{\operatorname{erfc}}

\begin{document}

\title{An exactly solvable macroscopic fluctuation theory of single-file diffusion}

\author{Sandeep Jangid~\orcidlink{0009-0002-3606-2580}}
\thanks{These authors contributed equally to this work.}
\affiliation{Department of Theoretical Physics, Tata Institute of Fundamental Research, Homi Bhabha Road, Mumbai 400005, India.}

\author{Soumyabrata Saha~\orcidlink{0009-0006-7539-8801}}
\thanks{These authors contributed equally to this work.}
\affiliation{Department of Theoretical Physics, Tata Institute of Fundamental Research, Homi Bhabha Road, Mumbai 400005, India.}

\author{Kapil Sharma~\orcidlink{0009-0005-9232-9616}}
\affiliation{Department of Theoretical Physics, Tata Institute of Fundamental Research, Homi Bhabha Road, Mumbai 400005, India.}

\author{Jitendra Kethepalli~\orcidlink{0000-0001-7326-0231}}
\affiliation{Laboratoire de Physique Th\'eorique et Mod\'elisation, CNRS UMR 8089, CY Cergy Paris Universit\'e, 95302 Cergy-Pontoise Cedex, France.}
\affiliation{JEIP, UAR 3573 CNRS, Collège de France, PSL Research University, 11 Place Marcelin Berthelot, 75321 Paris Cedex 05, France.}

\author{Benjamin Guiselin~\orcidlink{0000-0002-4663-2578}}
\affiliation{Laboratoire Charles Coulomb (L2C), Universit\'e de Montpellier, CNRS, Montpellier, France.}

\author{Jacopo De Nardis~\orcidlink{0000-0001-7326-0231}}
\affiliation{Laboratoire de Physique Th\'eorique et Mod\'elisation, CNRS UMR 8089, CY Cergy Paris Universit\'e, 95302 Cergy-Pontoise Cedex, France.}
\affiliation{JEIP, UAR 3573 CNRS, Collège de France, PSL Research University, 11 Place Marcelin Berthelot, 75321 Paris Cedex 05, France.}

\author{Tridib Sadhu~\orcidlink{0000-0003-0390-6257}}
\affiliation{Department of Theoretical Physics, Tata Institute of Fundamental Research, Homi Bhabha Road, Mumbai 400005, India.}

\date{\today}

\begin{abstract}
Single-file diffusion is a ubiquitous phenomenon in low-dimensional systems, arising in transport inside narrow channels. Its natural continuum model is a one-dimensional gas of extended Brownian hard rods (BHR). Perhaps owing to the perceived intractability of this problem, much of the literature has traditionally focused on lattice exclusion models, where integrability methods have yielded remarkable, albeit limited, exact results. A major recent advance comes from a formal solution of macroscopic fluctuation theory (MFT) for the exclusion process. Yet, despite the formal solution, only a handful of properties have been made explicit. We show that the corresponding MFT of the extended BHR gas is in fact exactly solvable through a canonical transformation. We demonstrate this by explicit computation of the large-deviation statistics of the tracer-position and integrated-current in both annealed and quenched ensembles. We further show that an analogous canonical transformation applies to the MFT of lattice gases with finite-volume exclusion, yielding corresponding tracer and current statistics. We validate our results using rare-event simulations for both the continuum and the lattice models.
\end{abstract}

\maketitle

\prlsection{Introduction} In the study of emergent collective behavior in non-equilibrium interacting many-particle systems, \emph{single-file diffusion}, in which particles undergo effectively one-dimensional diffusive motion without overtaking one another, has attracted sustained interest. This simple geometric constraint arises naturally in a variety of confined transport processes, including ion conduction through cell-membrane pores~\cite{1955_Hodgkin_The,1979_Urban_Ion}, molecular diffusion in porous media~\cite{1995_Gupta_Evidence,1996_Kulka_Nmr,1996_Hahn_Single,2007_Cheng_Observation,2014_Dvoyashkin_Single,2014_Pagliara_Channel}, the motion of colloidal assemblies in narrow channels~\cite{2000_Wei_Single,2004_Lutz_Single,2005_Lin_From,2016_Locatelli_Single}, and water transport in carbon nanotubes~\cite{2001_Hummer_Water,2002_Berezhkovskii_Single,2010_Das_Single}.

Single-file motion leads to non-trivial collective behavior such as strong spatiotemporal correlations~\cite{2015_Krapivsky_Dynamical,2023_Dandekar_Dynamical,2021_Poncet_Generalized}, slow memory relaxation~\cite{2013_Leibovich_Everlasting,2021_Poncet_Cumulant,2025_Poncet_Full}, anomalous diffusion~\cite{1965_Harris_Diffusion,1965_Jepsen_Dynamics,2003_Kollmann_Single}, and non-Gaussian fluctuations~\cite{2014_Krapivsky_Large,2014_Hegde_Universal,2017_Imamura_Large,2023_Grabsch_Driven}, none of which are typical of ordinary diffusion. These phenomena have been captured by a variety of low-dimensional stochastic particle models, ranging from exclusion processes~\cite{2007_Imamura_Dynamics,2017_Imamura_Large,2022_Grabsch_Exact} and random average processes~\cite{2001_Rajesh_Exact,2016_Cividini_Correlation,2024_Santra_Tracer} to Brownian particle gases~\cite{2008_Lizana_Single,2009_Barkai_Theory,2014_Krapivsky_Large,2014_Hegde_Universal,2023_Dandekar_Dynamical,2023_Touzo_Interacting}. Among them, exclusion processes, and in particular the symmetric simple exclusion process (SSEP), have been central to the theory of single-file diffusion. Their integrability has been instrumental, yielding exact results for the statistical properties of several key observables. The SSEP has also served as a paradigmatic test-bed for the hydrodynamic description of large-scale fluctuations, now widely known as macroscopic fluctuation theory (MFT)~\cite{2001_Bertini_Fluctuations,2005_Bertini_Current,2007_Derrida_Non,2015_Bertini_Macroscopic,2007_Tailleur_Mapping,2014_Krapivsky_Large,2017_Baek_Dynamical,2022_Mallick_Exact,2023_Agranov_Tricritical}.

Despite these successes, exact results for single-file diffusion, particularly for exclusion process, often rely on sophisticated microscopic techniques tailored to specific observables, initial ensembles, spatial settings, and exclusion rules~\cite{1994_Schutz_Non,2004_Schonherr_Exclusion,2008_Appert_Universal}. Consequently, progress in statistical characterization of collective properties has remained largely fragmented, with each new result~\cite{2001_Derrida_Free,2009_Derrida_Current,2012_Gorissen_Exact,2017_Imamura_Large,2021_Derrida_Large,2022_Grabsch_Exact} representing an independent technical achievement. Even the corresponding hydrodynamic theory of MFT has been solved \emph{explicitly} in only a handful of cases~\cite{2022_Mallick_Exact,2024_Saha_Large,2024_Grabsch_Semi,2025_Saha_Large,2026_Sharma_Large,2026_Saha_Effect}. This is further compounded by the intrinsically discrete nature of exclusion processes, which leaves them less directly connected to natural continuum realizations of single-file diffusion.

\begin{figure}[t]
\centering
\includegraphics[width=\linewidth, trim=0 20 0 35, clip]{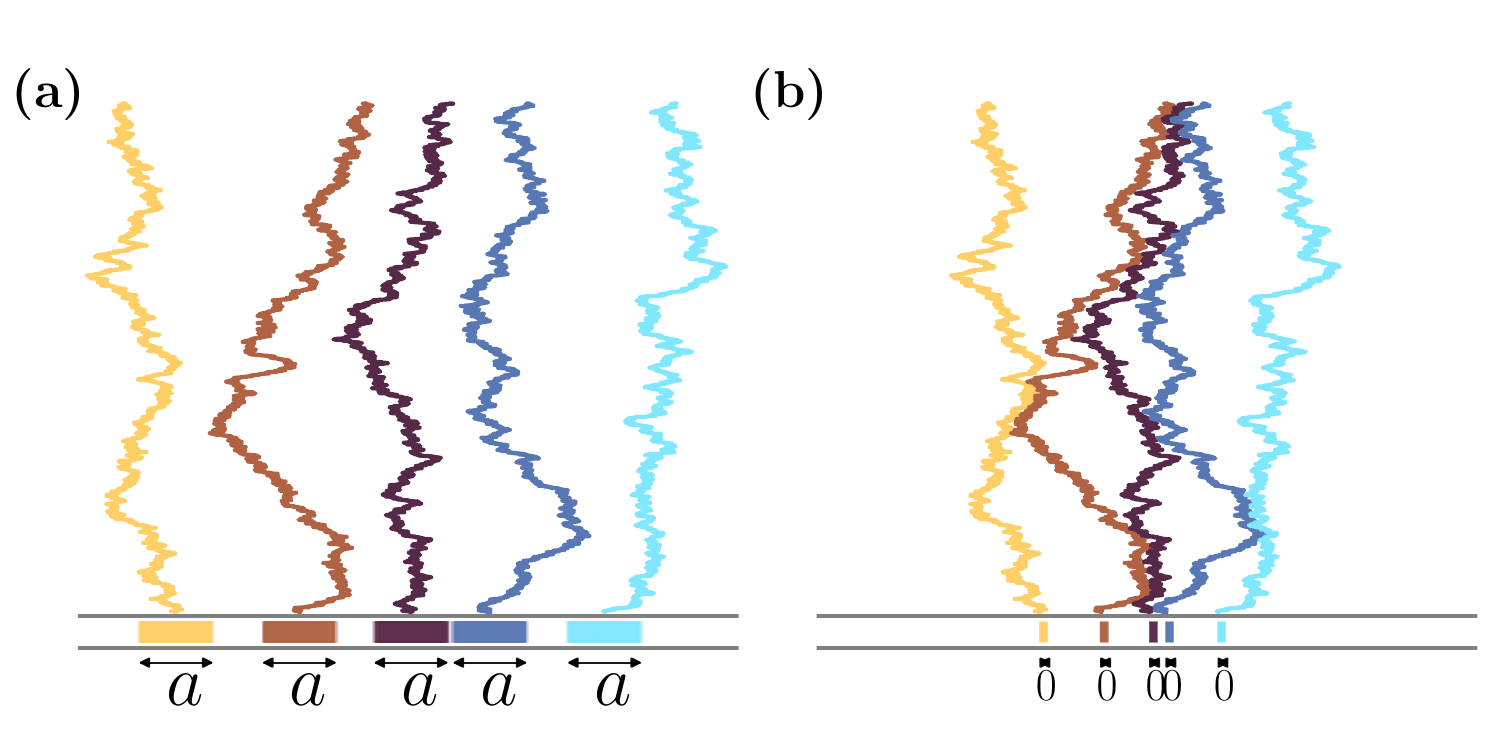}
\caption{Schematic trajectories of Brownian hard rods (a) of finite-size $a$ and (b) in the zero-size limit, $a\to 0$. The canonical transformation (see \eqref{eq:transform_mft}) maps (a) onto (b).}
\label{fig:schematic_trajectories}
\end{figure}

In this \emph{Letter}, we consider a gas of extended Brownian hard rods (BHR) (see Fig.~\ref{fig:schematic_trajectories}a) as a continuum model closer to natural single-file systems~\cite{1996_Hahn_Single,2000_Wei_Single,2004_Lutz_Single,2005_Lin_From,2007_Cheng_Observation,2010_Cambre_Experimental,2010_Das_Single,2014_Dvoyashkin_Single,2014_Pagliara_Channel,2016_Locatelli_Single}, and show that its hydrodynamic description is exactly solvable, yielding a number of explicit results, including some that have remained challenging~\cite{2021_Poncet_Cumulant,2025_Poncet_Full} for the exclusion process. The solution rests on a canonical transformation within the hydrodynamic theory, built on a duality between the BHR and non-interacting Brownians that we exploited in a recent work~\cite{2026_Saha_Universal}. This transformation explicitly solves the variational problem within MFT on the infinite line, and thus in principle offers full statistical characterization of generic macroscopic observables in BHR. We demonstrate it for the two typical observables in single-file diffusion, \emph{tracer-position} and \emph{integrated-current}, in the two conventional ensembles of initial states, \emph{annealed} and \emph{quenched}. Furthermore, we quantify the optimal trajectory along which the system realizes fluctuations of these observables. 

The BHR provides a rare example of a non-trivial MFT that is explicitly solvable, and arguably simpler than the few examples available from the exclusion paradigm~\cite{2022_Mallick_Exact, 2024_Mallick_Exact}. We further show that similar solutions of the MFT by a canonical transformation between elongated and point objects, extend beyond the continuum setting to lattice gases.

\prlsection{Model and setup} We consider a one-dimensional gas of Brownian hard rods of length $a$ on an infinite line subject to mutual hard-core repulsion prohibiting overlaps. Between collisions, each rod undergoes a free Brownian motion with bare diffusion coefficient $D_0=1$. Initially the system is prepared in a domain wall state where particles are placed without overlap following a Poisson distribution with an average step density profile, $\bar{\rho}_-$ and $\bar{\rho}_+$ to the left and right of the origin, respectively.

We focus on two key observables~\cite{2022_Banerjee_Role,2024_Grabsch_From,2025_Berlioz_Tracer,2025_Grabsch_Exact} to characterize the effect of single file constraint on the collective properties in BHR. (a) \emph{tracer-position} $z_t$, the position of the rod at time $t$ that started at the origin~\cite{1973_Levitt_Dynamics,2014_Krapivsky_Large,2014_Hegde_Universal,2015_Sadhu_Large,2017_Imamura_Large}, and (b) \emph{integrated-current} $Q_t$, the net number of rods that have crossed the origin from left to right in time $t$~\cite{2009_Derrida_Current,2009_Derrida_Current2,2012_Krapivksy_Fluctuations,2023_Dean_Effusion}. Their late-time statistics are quantified in terms of the corresponding cumulant-generating function.

Single-file systems are known to retain memory of their initial state~\cite{2013_Leibovich_Everlasting,2014_Lizana_Single,2026_Saha_Universal} even at late times. This is quantified by considering two different ensembles~\cite{2009_Derrida_Current2,2015_Krapivksy_Tagged,2015_Sadhu_Large,2023_Rana_Large} of initial fluctuations, following the terminology of disordered systems: the \emph{annealed ensemble}, denoted by the subscript $\mathcal{A}$, and the \emph{quenched ensemble}, denoted by the subscript $\mathcal{Q}$. For the two observables $z_t\equiv Q_t\equiv \mathcal{O}_t$, the cumulant-generating function scales as $\sqrt{t}$ with time $t$~\cite{2009_Derrida_Current2,2012_Krapivksy_Fluctuations,2015_Krapivksy_Tagged,2015_Sadhu_Large,2023_Rana_Large}, and the corresponding scaled cumulant-generating functions (SCGFs) in the two ensembles are defined as
\begin{equation}\label{eq:scgf_defn}
\mu_{\mathcal{A}}(\lambda)=\frac{1}{\sqrt{t}}\ln\overline{\big<\mathrm{e}^{\lambda\mathcal{O}_t}\big>}\,,\,\,\mu_{\mathcal{Q}}(\lambda)=\frac{1}{\sqrt{t}}\overline{\ln\big<\mathrm{e}^{\lambda\mathcal{O}_t}\big>},
\end{equation}
where the angular brackets denote an average over the stochastic evolution for fixed initial positions, while the overline denotes an average over the initial positions of the rods.

\begin{figure*}[t]
\centering
\includegraphics[width=\linewidth, trim=90 16 45 30, clip]{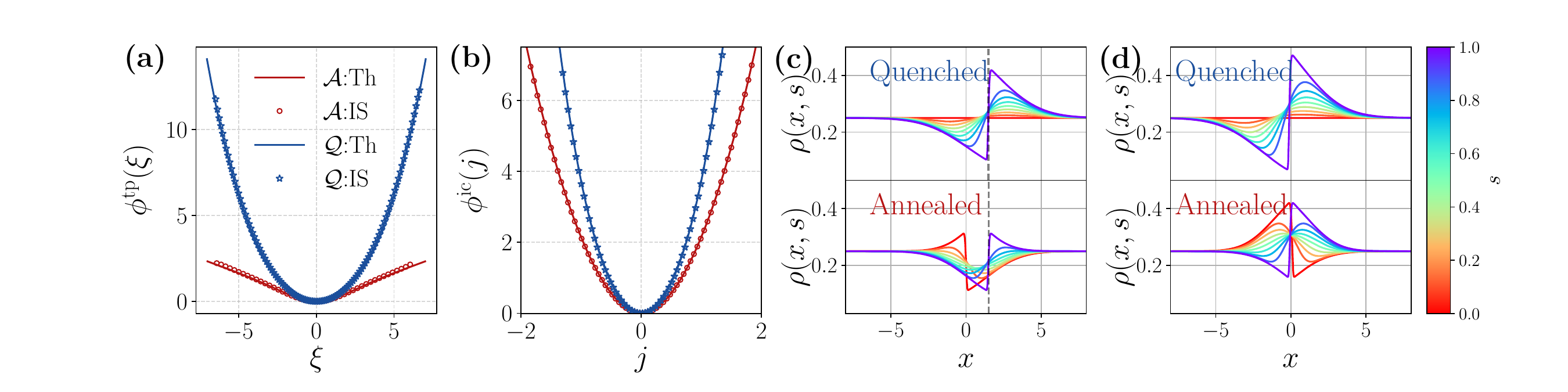}
\caption{(a,b) LDFs for (a) tracer-position and (b) integrated-current, defined by $\phi(\mathfrak{o})=-\tfrac{1}{\sqrt{t}}\ln\mathcal{P}\left(\tfrac{\mathcal{O}_t}{\sqrt{t}}=\mathfrak{o}\right)$ with $\mathfrak{o}=\xi,\,j$, respectively. Analytical predictions \eqref{eq:scgf_bhr}, shown as solid lines, are compared with importance-sampling simulations~\cite{2014_Hartmann_High,2026_Yerrababu_Dynamical,2026_Saha_Universal}, shown as symbols. Blue and red denote the quenched and annealed ensembles, respectively. (c,d) Optimal density trajectories associated with the prescribed values (c) $\xi=1.5$ for the tracer-position and (d) $j=0.375$ for the integrated-current. The upper and lower panels correspond to the quenched and annealed ensembles, respectively. In the annealed ensemble, the optimal trajectories obey the time-reversal symmetries $\rho(x,s)=\rho(\xi-x,1-s)$ for the tracer-position and $\rho(x,s)=\rho(-x,1-s)$ for the integrated-current. Parameters are set to $a=1$ and $\bar{\rho}(x)=0.25$; simulations are performed with $N=16385$ particles.}
\label{fig:ldf_traj}
\end{figure*}

\prlsection{Results} The SCGFs of the tracer-position (denoted by superscript `tp') and of the integrated-current (denoted by superscript `ic') are given by
\begin{subequations}\label{eq:scgf_bhr}
\begin{align}
\mu^{\rm tp}(\lambda)&=\xi\lambda+\chi(\xi,B),\label{eq:tracer_scgf}\\
\mu^{\rm ic}(\lambda)&=\xi B+\chi(\xi,\lambda-aB),\label{eq:current_scgf}
\end{align}
where the auxiliary variables $\xi$ and $B$ are fixed by the saddle point conditions $\partial_\xi\mu=\partial_B\mu=0$. The ensemble dependence only enters through the function $\chi$, which corresponds to the SCGF of the \emph{local-height} for zero-length BHR with renormalized step-densities $\bar{r}_\mp=\tfrac{\bar{\rho}_\mp}{1-a\bar{\rho}_\mp}$~\cite{Supp_Mat}. For the annealed and quenched ensembles, respectively,
\begin{align}
\!\!\!\!\chi_{\mathcal{A}}(\xi,B)&=\!\sum_{\varepsilon=(\pm)}\bar{r}_{-\varepsilon}(\mathrm{e}^{\varepsilon B}-1)\int_{\varepsilon\xi}^{\infty}\!\mathrm{d}u\,H(u),\label{eq:chi_bhr_ann}\\
\!\!\!\!\chi_{\mathcal{Q}}(\xi,B)&=\!\sum_{\varepsilon=(\pm)}\bar{r}_{-\varepsilon}\int_{\varepsilon\xi}^{\infty}\!\mathrm{d}u\ln\big[1+(\mathrm{e}^{\varepsilon B}-1)H(u)\big],\label{eq:chi_bhr_que}
\end{align}
\end{subequations}
where, $H(u)=\tfrac{1}{2}\E\tfrac{u}{2}$. The local-height $h_t^{(\mathcal{Z})}$ at time $t$ and position $\mathcal{Z}$ measures the excess of the integrated-current through the origin up to time $t$ over the number of particles contained between $0$ and $\mathcal{Z}$ at time $t$~\cite{1997_Schutz_Duality,2017_Imamura_Large}. The strikingly similar structures~\eqref{eq:scgf_bhr} of the tracer-position and integrated-current SCGFs reflect their common connection to this local-height observable~\cite{2024_Mallick_Exact,Supp_Mat}. In Figs.~\ref{fig:ldf_traj}a and~\ref{fig:ldf_traj}b, we provide numerical confirmation of the large-deviation functions (LDFs) for the tracer-position and integrated-current, respectively, obtained by Legendre transforming the corresponding SCGFs~\eqref{eq:scgf_bhr} for a uniform initial profile.

Several consequences and connections follow from the exact SCGFs in \eqref{eq:scgf_bhr}. The two observables exhibit qualitatively different dependence on the hard-rod length $a$. While the tracer-position SCGF depends on $a$ only through the renormalized densities $\bar{r}_\mp$, the integrated-current SCGF retains an additional explicit dependence on $a$. These results extend previous studies~\cite{2023_Rizkallah_Duality,2024_Grabsch_From,2026_Saha_Universal}, which were largely restricted to equilibrium settings, to inhomogeneous initial conditions. They further reveal a fluctuation symmetry~\cite{1993_Evans_Probability,1995_Gallavotti_Dynamical,1998_Kurchan_Fluctuation,1999_Lebowitz_A} for an interacting system in continuum. In the annealed ensemble, the SCGFs satisfy~\cite{Supp_Mat}
\begin{equation}\label{eq:gcems_symm}
\mu_{\mathcal{A}}(\lambda)=\mu_{\mathcal{A}}(c_n-\lambda)\,,\,\,c_n=\int_{\bar{\rho}_-}^{\bar{\rho}_+}\mathrm{d}w\,w^n\frac{2D(w)}{\sigma(w)}
\end{equation}
where $n=1\,(0)$ for the tracer-position (integrated-current), and 
\begin{equation}\label{eq:transport_bhr}
D(\rho)=\frac{1}{(1-a\rho)^2}\,,\,\,\sigma(\rho)=2\rho
\end{equation}
denote the transport coefficients, diffusivity and mobility, respectively, of the BHR~\cite{1991_Spohn_Large,2005_Lin_From,2026_Grabsch_Macroscopic,2026_Saha_Bottom}. Remarkably, the annealed integrated-current SCGF coincides exactly with its counterpart for Hamiltonian hard-rods~\cite{2026_Kethepalli_Ballistic}, where particles move ballistically between collisions and exchange velocities upon contact. This extends the recently observed universality between these two distinct dynamical settings~\cite{2026_Saha_Universal} to annealed integrated-current statistics.

Both the tracer-position and integrated-current have Gaussian typical fluctuations, with variances obeying the annealed--quenched relation~\cite{2012_Krapivksy_Fluctuations,2013_Leibovich_Everlasting,2015_Krapivksy_Tagged} $\langle\mathcal{O}_t^2\rangle_{\mathcal{A}}=\sqrt{2}\langle\mathcal{O}_t^2\rangle_{\mathcal{Q}}$. Their far tails, however, differ more distinctly. From the large-$\lambda$ behavior of the SCGFs, we find
$\mu_{\mathcal{A}/\mathcal{Q}}^{\rm ic}(\lambda)$ and $\mu_{\mathcal{Q}}^{\rm tp}(\lambda)$
to scale as $\lambda^{3/2}$~\cite{2009_Derrida_Current2,2014_Meerson_Extreme,2015_Krapivksy_Tagged}, while $\mu_{\mathcal{A}}^{\rm tp}(\lambda)$ has a finite support,
$-\bar{r}_-\leq\lambda\leq\bar{r}_+$. Thus the integrated-current LDFs in both ensembles, as well as the quenched tracer LDF, have cubic far tails, whereas the annealed tracer LDF has a linear tail, reflecting quasi-stationarity of tracer-position distribution at large $z_t$~\cite{2020_Barkai_Packets}.

\prlsection{Optimal trajectory} The optimal trajectory is the most probable evolution of the coarse-grained density profile $\rho(x,s)$ in rescaled length $x$ and time $s$ with $\sqrt{t}$ and $t$, respectively, that drives the system towards a targeted value of the observable at time $t$. For the BHR, this optimal trajectory is expressed parametrically in terms of a renormalized density field $r(X,s)$ through $\rho(x,s)=\tfrac{r(X,s)}{1+ar(X,s)}$ with $x=X+a\int_{\mathfrak{z}(s)}^{\infty}\mathrm{d}X'r(X',s)$, where $\mathfrak{z}(s)$ denotes the rescaled tracer-position at rescaled time $s$. In the annealed and quenched ensembles, $r(X,s)$ is given by~\cite{Supp_Mat}
\begin{subequations}\label{eq:optimal_traj_bhr}
\begin{align}
r_{\mathcal{A}}(X,s)&=\mathcal{R}_s(X)\int_{-\infty}^{\infty}\mathrm{d}X'\mathcal{G}_s(X-X')\frac{\bar{r}(X')}{\mathcal{V}_0(X')},\label{eq:r_ann_opt}\\
r_{\mathcal{Q}}(X,s)&=\mathcal{R}_s(X)\int_{-\infty}^{\infty}\mathrm{d}X'\mathcal{G}_s(X-X')\frac{\bar{r}(X')}{\mathcal{R}_0(X')},\label{eq:r_que_opt}\\
\mathcal{R}_s(X)&=\int_{-\infty}^{\infty}\mathrm{d}X'\mathcal{G}_{1-s}(X-X')\mathcal{V}_1(X').\label{eq:R_cal}
\end{align}
Here, $\mathcal{G}_s(X)=(4\pi s)^{-\frac{1}{2}}\mathrm{e}^{-\frac{X^2}{4s}}$ is the Green's function, and $\bar{r}(X)=\bar{r}_-\theta(-X)+\bar{r}_+\theta(X)$ denotes the initial average step-density profile. The functions $\mathcal{V}_{0}(X)$ and $\mathcal{V}_{1}(X)$ encode the observable dependence. For the tracer-position, they are given by
\begin{equation}
\mathcal{V}_0(X)=\mathrm{e}^{\frac{\lambda\theta(X)}{r(\xi,1)}}\,,\,\,\mathcal{V}_1(X)=\mathrm{e}^{\frac{\lambda\theta(X-\xi)}{r(\xi,1)}}
\end{equation}
whereas, for the integrated-current
\begin{equation}
\mathcal{V}_0(X)=\mathrm{e}^{\frac{\lambda\theta(X)}{1+ar(aj,1)}}\,,\,\,\mathcal{V}_1(X)=\mathrm{e}^{\frac{\lambda\theta(X-aj)}{1+ar(aj,1)}}.
\end{equation}
\end{subequations}
Here, $\xi=\tfrac{z_t}{\sqrt{t}}$ and $j=\tfrac{Q_t}{\sqrt{t}}$ denote the rescaled values of the tracer-position and integrated-current, respectively.

The optimal density profile~\eqref{eq:optimal_traj_bhr} at successive stages of the evolution is shown in Figs.~(\ref{fig:ldf_traj}c, \ref{fig:ldf_traj}d) for a given value of the tracer-position and integrated-current, respectively, starting from a homogeneous density profile. This full characterization is striking considering that in SSEP~\cite{2022_Mallick_Exact} and related integrable models~\cite{2021_Krajenbrink_Inverse,2022_Bettelheim_Inverse,2024_Bettelheim_Complete,2026_Krajenbrink_Integrability} optimal profile could be made explicit only at the initial and final times.

In the remainder of this \emph{Letter}, we show how these results are obtained within the hydrodynamic description of the MFT.

\prlsection{MFT formalism} At macroscopic length and time scales, BHR is described by a coarse-grained density $\rho(x,s)$ whose evolution is governed by the fluctuating hydrodynamic equation~\cite{2026_Saha_Bottom}
\begin{equation}\label{eq:fhd_eqn}
\partial_s\rho=\partial_x\bigg(D(\rho)\partial_x\rho+\frac{\sqrt{\sigma(\rho)}}{t^{1/4}}\eta\bigg),
\end{equation}
with the transport coefficients in \eqref{eq:transport_bhr} and the large observation time $t$ setting the diffusive (macroscopic) length scale $\sqrt{t}$. Here, $\eta(x,s)$ is a zero-mean Gaussian white noise with covariance $\left<\eta(x,s)\eta(x',s')\right>=\delta(x-x')\delta(s-s')$.

For a fixed initial profile $\rho(x,0)$, the fluctuating hydrodynamics \eqref{eq:fhd_eqn} assigns a dynamical cost to each density evolution. The Martin–Siggia–Rose–Janssen–De-Dominicis (MSRJD) formalism~\cite{1973_Martin_Statistical,1976_Janssen_On,1976_Dominicis_Techniques,1978_Dominicis_Field} captures this cost through the dynamical action
\begin{equation}\label{eq:mft_action}
\!\!\!\mathcal{S}_a[\rho,\hat{\rho}]\!=\!\!\int_{0}^{1}\!\!\mathrm{d}s\!\!\int\!\!\mathrm{d}x\bigg\{\!\hat{\rho}\partial_s\rho-\bigg[\frac{\sigma(\rho)}{2}\partial_x\hat{\rho}-D(\rho)\partial_x\rho\bigg]\partial_x\hat{\rho}\!\bigg\}
\end{equation}
where $\hat{\rho}(x,s)$ is the response field conjugate to the density, arising from local particle conservation. The subscript $a$ refers to the length of the rods (see Fig.~\ref{fig:schematic_trajectories}).

For the domain-wall initial state, the density profile $\rho(x,0)$ fluctuates around the average $\bar{\rho}(x)$ with the probability $\mathcal{P}[\rho(x,0)]\asymp\mathrm{e}^{-\sqrt{t}\,\mathcal{F}_a[\rho(x,0)]}$, where the free-energy cost~\cite{2007_Derrida_Non} of such deviations is
\begin{equation}\label{eq:initial_free_energy}
\!\!\mathcal{F}_a[\rho(x,0)]=\int_{-\infty}^{\infty}\!\mathrm{d}x\int_{\bar{\rho}(x)}^{\rho(x,0)}\!\mathrm{d}w\,(\rho(x,0)-w)\frac{2D(w)}{\sigma(w)}.
\end{equation}
Together with the action \eqref{eq:mft_action}, this free-energy functional defines the probability weight over density trajectories, from which the statistics of arbitrary observables $\mathcal{O}_t$ expressible as functionals of these trajectories can be obtained.

Both observables considered in this \emph{Letter} obey the scaling $\mathcal{O}_t\simeq\sqrt{t}\,\mathfrak{o}[\rho]$ for large $t$. The tracer-position is $z_t\simeq\sqrt{t}\,\xi[\rho]$, with $\xi\equiv\mathfrak{o}$ defined~\cite{2014_Krapivsky_Large,2015_Krapivksy_Tagged} in terms of density $\rho(x,s)$ as 
\begin{equation}\label{eq:z_t_defn}
\int_{0}^{\xi}\mathrm{d}x\,\rho(x,1)=\int_{0}^{\infty}\mathrm{d}x\big(\rho(x,1)-\rho(x,0)\big),
\end{equation}
using the single-file constraint, while the integrated-current is $Q_t=\sqrt{t}\,j[\rho]$ with $j\equiv\mathfrak{o}$ taking the explicit form~\cite{2009_Derrida_Current2,2012_Krapivksy_Fluctuations}
\begin{equation}\label{eq:Q_t_defn}
j=\int_{0}^{\infty}\mathrm{d}x\big(\rho(x,1)-\rho(x,0)\big).
\end{equation}
Within MFT, the SCGF for these observables in annealed ensemble defined in \eqref{eq:scgf_defn}, is then expressed as a path-integral
\begin{equation}\label{eq:path_integral_observable}
\mu_{\mathcal{A}}(\lambda)=\frac{1}{\sqrt{t}}\ln\!\!\int\mathcal{D}[\rho,\hat{\rho}]\mathrm{e}^{\sqrt{t}\left(\lambda\mathfrak{o}[\rho]-\mathcal{F}_a[\rho(x,0)]-\mathcal{S}_a[\rho,\hat{\rho}]\right)},
\end{equation}
where $\mathcal{S}_a$ and $\mathcal{F}_a$ are given in~\eqref{eq:mft_action} and~\eqref{eq:initial_free_energy}, respectively. For the SCGF~\eqref{eq:scgf_defn} in the quenched ensemble, the smooth logarithmic function inside the average over initial positions effectively selects contributions only from around the average initial profile $\bar{\rho}(x)$. Consequently, $\mu_{\mathcal{Q}}(\lambda)$ is computed using a similar path-integral in \eqref{eq:path_integral_observable}, with $\rho(x,0)=\bar{\rho}(x)$, and setting $\mathcal{F}_a[\rho(x,0)]=0$.

Since the exponent in \eqref{eq:path_integral_observable} is extensive in $\sqrt{t}$, the large-$t$ asymptotics of the generating-function is governed by a saddle point. The corresponding saddle-point equations, often referred to as the MFT equations, determine the \emph{optimal trajectories} of the density and response fields, and evaluating the exponent along these trajectories yields the \emph{SCGF} of the observable. The MFT equations for the BHR are
\begin{subequations}\label{eq:mft_eqn_hard_rod}
\begin{align}
\partial_t\rho&=\partial_x\big(D(\rho)\partial_x\rho-\sigma(\rho)\partial_x\hat{\rho}\big),\label{eq:rho_el_eqn}\\
\partial_t\hat{\rho}&=-D(\rho)\partial_x^2\hat{\rho}-\frac{\sigma'(\rho)}{2}(\partial_x\hat{\rho})^2,\label{eq:rhohat_el_eqn}
\end{align}
with temporal boundary conditions specific to the observable. For annealed ensemble
\begin{align}
\hat{\rho}(x,0)&=-\lambda\frac{\delta\mathfrak{o}[\rho]}{\delta\rho(x,0)}+\int_{\bar{\rho}(x)}^{\rho(x,0)}\mathrm{d}w\frac{2D(w)}{\sigma(w)},\label{eq:initial_cond}\\
\hat{\rho}(x,1)&=\lambda\frac{\delta\mathfrak{o}[\rho]}{\delta\rho(x,1)},\label{eq:final_cond}
\end{align}
\end{subequations}
while in the quenched ensemble, the initial-time condition in \eqref{eq:initial_cond} is replaced by $\rho(x,0)=\bar{\rho}(x)$.

\prlsection{Point-particle mapping} Solving these nonlinear coupled MFT equations (\ref{eq:rho_el_eqn}-\ref{eq:rhohat_el_eqn}) with the observable-dependent boundary conditions (\ref{eq:initial_cond}-\ref{eq:final_cond}) is a challenging task. Exact solutions are known for only a few models and are typically restricted to specific observables and ensembles, such as the annealed tracer-position and integrated-current for the SSEP~\cite{2022_Mallick_Exact,2024_Mallick_Exact}. Our results in (\ref{eq:scgf_bhr},\ref{eq:optimal_traj_bhr}) therefore represent a significant advance in this direction. The key to solving this nontrivial MFT problem is a change of coordinates $x\to X$, combined with a canonical transformation of the density and response fields, $(\rho,\hat{\rho})\to(r,\hat{r})$, defined by~\cite{Supp_Mat}
\begin{subequations}\label{eq:transform_mft}
\begin{align}
X&=x-a\int_{\mathfrak{z}(s)}^{x}\mathrm{d}x'\rho(x',s),\label{eq:transform_coordinate}\\
r(X,s)&=\frac{\rho(x,s)}{1-a\rho(x,s)},\label{eq:transform_density}\\
\hat{r}(X,s)&=\hat{\rho}(x,s)-a\int_{\mathfrak{z}(s)}^{x}\mathrm{d}x'\rho(x',s)\partial_{x'}\hat{\rho}(x',s),\label{eq:transform_response}
\end{align}
\end{subequations}
where $\mathfrak{z}(s)$ denotes the rescaled position at rescaled time $s$ of the particle initially located at the origin. In the transformed variables, the problem is exactly solvable~\cite{Supp_Mat}, allowing us to determine explicitly the SCGFs and corresponding optimal trajectories for the tracer-position and integrated-current in both the annealed and quenched ensembles. A similar reduction applies for the quenched case.

The solvability of the MFT for BHR becomes transparent a posteriori from the structure of the canonical transformation~\eqref{eq:transform_mft}. Under the coordinate transformation~\eqref{eq:transform_coordinate}, the BHR is mapped onto a point-particle (PP) system, with the underlying microscopic correspondence illustrated schematically in Fig.~\ref{fig:schematic_trajectories}. The mapped system is described by the renormalized density field $r(X,s)$~\eqref{eq:transform_density}, with transport coefficients $D(r)=1$ and $\sigma(r)=2r$, corresponding to the vanishing rod-length limit ($a\to 0$) of~\eqref{eq:transport_bhr}. The observables transform accordingly as $\mathcal{O}_t[\rho]=\widetilde{\mathcal{O}}_t[r]$ and consequently, the annealed generating-function in~\eqref{eq:scgf_defn} can be evaluated using $\overline{\big<\mathrm{e}^{\lambda\mathcal{O}_t}\big>}_a=\overline{\big<\mathrm{e}^{\lambda\widetilde{\mathcal{O}}_t}\big>}_{a\to 0}$ from point-particle limit. 

Interestingly, the functional dependence of tracer-position is invariant~\cite{Supp_Mat} under the mapping, $z_t[\rho]\mapsto\widetilde{z}_t[r]= z_t[r]$ and the transformation \eqref{eq:transform_mft} only replaces the density field $\rho$ by $r$, immediately leading to the scgf \eqref{eq:tracer_scgf} utilizing earlier PP-results~\cite{2009_Derrida_Current2}. By contrast, the integrated-current maps non-trivially to a functional of $r$, $Q_t[\rho]\mapsto\widetilde{Q}_t[r]$, which corresponds~\cite{Supp_Mat} to the local-height in PP-coordinates, $\widetilde{Q}_t[r]\equiv h_t^{(\mathcal{Z})}[r]=Q_t[r]-\sqrt{t}\int_{0}^{\mathcal{Z}/\sqrt{t}}\mathrm{d}X\,r(X,1)$, evaluated at the observable-dependent position $\mathcal{Z}=a\widetilde{Q}_t[r]$, yielding the non-trivial scgf \eqref{eq:current_scgf}.

The coordinate-transformation~\eqref{eq:transform_coordinate} was proposed~\cite{2024_Doyon_New,2026_Kethepalli_Ballistic,2026_Doyon_Towards,2026_Urilyon_Simulating} in the context of integrable Hamiltonian systems to effectively removing the scattering shift and map the interacting system onto a PP-system. Such mappings provide a powerful route to macroscopic properties that would otherwise be difficult to determine explicitly. In the \emph{End Matter}, we show that the canonical-transformation approach developed for the BHR in this \emph{Letter} extends naturally to single-file diffusion on a lattice with finite-range spatial exclusion~\cite{1968_Macdonald_Kinetics,1969_Macdonald_Concerning,2003_Lakatos_Totally,2003_Shaw_Totally,2004_Schonherr_Exclusion,2005_Schonherr_Hard,2011_Gupta_Driven,2013_Krapivsky_Dynamics,2026_Saha_Bottom}.

\prlsection{Conclusion} In this \emph{Letter}, we provide full statistical description of two central observables in the BHR: the tracer-position and integrated-current. Within the framework of MFT, we explicitly compute their SCGFs in both the annealed and quenched ensembles, along with the optimal trajectories that realize the corresponding fluctuations. The key step is a canonical transformation of the hydrodynamic fields, which renders this interacting continuum model exactly solvable, thus providing a significant advance in exact solutions of MFT. Non-triviality of this MFT is particularly evident from the long-range correlations, similar to SSEP, apparent~\cite{2009_Bertini_Towards,2016_Sadhu_Correlations} from the density dependence of the ratio $\tfrac{\sigma'(\rho)}{D(\rho)}$. This level of explicit solvability places the BHR at a distinct advantage over traditional exclusion models, for which comparable solutions remain largely unavailable. Moreover, by retaining continuum space and finite particle extent, BHR closely reflects physically realizable single-file systems, making our findings directly relevant to experiments~\cite{1996_Hahn_Single,2000_Wei_Single,2004_Lutz_Single,2005_Lin_From,2007_Cheng_Observation,2010_Cambre_Experimental,2010_Das_Single,2014_Dvoyashkin_Single,2014_Pagliara_Channel,2016_Locatelli_Single}. In fact, BHR is a limiting description~\cite{1953_Rouse_Theory,2014_Lips_Brownian,2025_Yuste_Single,2026_Grabsch_Macroscopic} of dense Brownian gases with sharply varying repulsive interactions and a finite-core. 

Our work opens several exciting and timely directions for future study. Extending the results obtained here to a broad range of initial conditions, observables, geometries, and multi-time, conditional statistics~\cite{2007_Derrida_Non,2008_Appert_Universal,2015_Krapivsky_Dynamical,2015_Krapivsky_Melting,2019_Derrida_Large,2025_Derrida_Lecture,2025_Poncet_Full} would provide a more comprehensive quantitative characterization of macroscopic fluctuations in non-equilibrium transport, thereby building a tractable counterpart to the phenomenology of exclusion processes. Along this direction, exploring the emergence of KPZ-scaling in the Brownian setting, parallel to the ASEP~\cite{2007_Derrida_Non,2015_Mallick_The}, may reveal new connections between single-file transport and fluctuating interfaces~\cite{1991_Majumdar_Tag,1997_Krug_Persistence}.

\prlsection{Acknowledgments} This research is supported in part by the International Centre for Theoretical Sciences (ICTS) through participation in the program \emph{Hydrodynamics, Fluctuations, and Noise in Quantum and Classical Systems 2025} (code: ICTS/hydrodynamics2025/12). JDN and JK are funded by the ERC Starting Grant 101042293 (HEPIQ) and the ANR-22-CPJ1-0021-01. TS, SS, SJ and KS acknowledge financial support from the Department of Atomic Energy, Government of India, under Project Identification Number RTI-4012. We thank the Department of Theoretical Physics, TIFR, Mumbai, for providing computational facilities, and Ajay Salve and Kapil Ghadiali for computational support. TS further acknowledges support from the International Research Project (IRP) ``Classical and Quantum Dynamics in Out-of-Equilibrium Systems,'' funded by CNRS, France.

\bibliographystyle{apsrev4-2}
\bibliography{reference}

\renewcommand{\theequation}{E.\arabic{equation}}
\setcounter{equation}{0}

\renewcommand{\thefigure}{E.\arabic{figure}}
\setcounter{figure}{0}

\raggedbottom

\addtolength{\abovedisplayskip}{-1pt}
\addtolength{\belowdisplayskip}{-1pt}

\par\bigskip
{\centering\large\bfseries End Matter\par}
\medskip

\par\medskip
{\centering\bfseries\itshape Local-height statistics\par}
\smallskip

In the \emph{Letter}, we analyzed the tracer-position and integrated-current. Here, we extend our results to the local-height observable for both ensembles of the BHR that appeared in the context of \eqref{eq:scgf_bhr}. The local-height $h_t^{(\mathfrak{z})}=\sqrt{t}\,\kappa$, measured at $\mathfrak{z}=\sqrt{t}\,\zeta$, is defined as
\begin{equation}
\kappa=j-\int_{0}^{\zeta}\mathrm{d}x\,\rho(x,1)
\end{equation}
Its SCGF for both ensembles is given by
\begin{equation}
\mu^{\rm lh}(\zeta,\lambda)=(\xi-\zeta)B+\chi(\xi,\lambda-aB),
\end{equation}
where $\xi$ and $B$ are determined from the conditions $\partial_\xi\mu=\partial_B\mu=0$. Details about the ensemble enters through $\chi$ whose expression is in \eqref{eq:chi_bhr_ann} and \eqref{eq:chi_bhr_que} for the annealed and quenched ensembles, respectively, and it corresponds to local-height for a PP-system described by the renormalized density field $r(X,s)$.

For both ensembles, the large-$|\lambda|$ asymptotics of the SCGF is given by $\mu^{\rm lh}(\lambda)\sim|\lambda|^{3/2}$. The SCGF in annealed ensemble further exhibits the fluctuation symmetry
\begin{equation}
\mu_\mathcal{A}^{\rm lh}(\xi,\lambda)=\mu_\mathcal{A}^{\rm lh}(-\xi,c_0-\lambda)-c_1\xi,
\end{equation}
with $c_{0,1}$ as defined in \eqref{eq:gcems_symm}.

For the homogeneous initial average profile $\bar{\rho}(x)=\bar{\rho}$, the average local-height is given by
\begin{equation}
\big<h_t^{(\mathfrak{z})}\big>\simeq-\bar{\rho}\,\mathfrak{z}
\end{equation}
in both ensembles. The variance depends on the ensemble and reads
\begin{align}
\frac{\Big<\big[h_t^{(\mathfrak{z})}\big]^2\Big>_{\mathcal{A}}}{\sqrt{t}}&\simeq\frac{2\bar{\rho}(1-a\bar{\rho})}{\sqrt{\pi}}\equiv\frac{\big<Q_t^2\big>_{\mathcal{A}}}{\sqrt{t}},\\
\frac{\Big<\big[h_t^{(\mathfrak{z})}\big]^2\Big>_{\mathcal{Q}}}{\sqrt{t}}&\simeq\frac{\sqrt{2}\bar{\rho}(1-a\bar{\rho})}{\sqrt{\pi}}\equiv\frac{\big<Q_t^2\big>_{\mathcal{Q}}}{\sqrt{t}}.
\end{align}

The corresponding optimal trajectories leading to a specific value of the local-height are given by (\ref{eq:r_ann_opt}-\ref{eq:R_cal}), with the functions $\mathcal{V}_{0,1}$ now taking the form
\begin{equation}
\mathcal{V}_0(X)=\mathrm{e}^{\frac{\lambda\theta(X)}{1+ar(\zeta+a\kappa,1)}}\,,\,\,\mathcal{V}_1(X)=\mathrm{e}^{\frac{\lambda\theta(X-\zeta-a\kappa)}{1+ar(\zeta+a\kappa,1)}}.
\end{equation}

\textit{Relation to tracer-position and integrated-current:} The tracer-position, defined implicitly in \eqref{eq:z_t_defn}, corresponds to the zero of the local-height field,
\begin{equation}
z_t[\rho]=\mathfrak{z}\Leftrightarrow h_t^{(\mathfrak{z})}=0.
\end{equation}
By contrast, the integrated-current, defined in \eqref{eq:Q_t_defn}, is obtained directly as the local-height at the origin,
\begin{equation}
Q_t[\rho]\equiv h_t^{(0)}[\rho].
\end{equation}

\par\medskip
{\centering\bfseries\itshape Non-equilibrium initial conditions\par}
\smallskip

\begin{figure}[t]
\centering
\includegraphics[width=\linewidth, trim=10 15 0 0, clip]{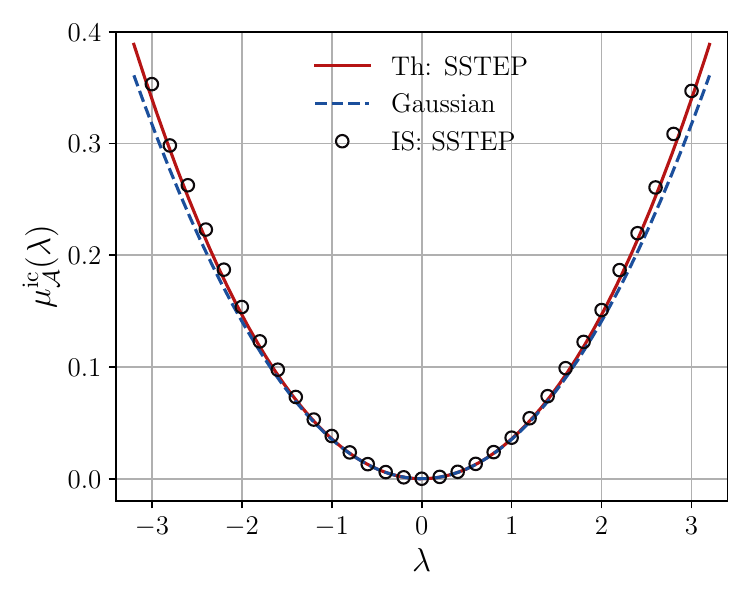}
\caption{Annealed integrated-current SCGF for the triple-site symmetric simple exclusion process (SSTEP) where particles occupy an additional $M=2$ sites on each side of its current site. The solid line indicates the analytical expression (\ref{eq:current_scgf}, \ref{eq:psi_multi_ssep_anneal}) while the black dots denote numerical results obtained from rare-event simulations based on cloning algorithm~\cite{2006_Giardina_Direct,2007_Lecomte_Numerical} (parameters used: $\bar{\rho}=\tfrac{1}{12}$, $t=170$ and number of clones$=10^5$). The dashed line denotes Gaussian approximation, revealing the non-Gaussianity of the fluctuations.}
\label{fig:triple_ssep_anneal_current}
\end{figure}

The mean tracer-position and integrated-current take the form
\begin{equation}
\big<\mathcal{O}_t\big>\sim\sqrt{t}\,\overline{\mathfrak{o}},
\end{equation}
where $\overline{\mathfrak{o}}=\overline{\xi}$ and $\overline{j}$ for the tracer-position and integrated-current, respectively, given by the unique solutions of the integral equations
\begin{align}
\overline{\xi}&=\frac{\bar{r}_--\bar{r}_+}{\bar{r}_+}\int_{\overline{\xi}}^{\infty}\mathrm{d}u\,H(u),\\
\overline{j}&=\frac{\bar{r}_--\bar{r}_+}{1+a\bar{r}_+}\int_{a\overline{j}}^{\infty}\mathrm{d}u\,H(u).
\end{align}
The average values are thus independent of the choice of ensemble, with the dependence first appearing at the level of fluctuations. The variances of tracer-position and integrated-current are respectively given by
\begin{equation}
\frac{\big<z_t^2\big>}{\sqrt{t}}=\frac{K(\overline{\xi})}{[L(\overline{\xi})]^2}\,,\,\,\frac{\big<Q_t^2\big>}{\sqrt{t}}=\frac{K(a\overline{j})}{[1+aL(a\overline{j})]^2},
\end{equation}
where
\begin{align}
K_\mathcal{A}(x)&=\sum_{\varepsilon=(\pm)}\bar{r}_{-\varepsilon}\int_{\varepsilon x}^{\infty}\mathrm{d}u\,H(u)\\
K_\mathcal{Q}(x)&=\sum_{\varepsilon=(\pm)}\bar{r}_{-\varepsilon}\int_{\varepsilon x}^{\infty}\mathrm{d}u\,H(u)H(-u)\\
L(x)&=\sum_{\varepsilon=(\pm)}\bar{r}_{-\varepsilon}H(\varepsilon x)
\end{align}

A particularly interesting limit arises for $\bar{\rho}_+=0$, where the tracer-position exhibits Gumbel statistics. Its mean and variance have the asymptotic forms
\begin{equation}
\big<z_t\big>\simeq\sqrt{2t\ln t}\,,\,\,\big<z_t^2\big>\simeq\frac{\pi^2t}{3\ln t}.
\end{equation}
Remarkably, these extreme-value statistics are ensemble independent and insensitive to the details of interactions~\cite{2007_Sabhapandit_Statistical,2017_Imamura_Large}.

\par\medskip
{\centering\bfseries\itshape The multi-site exclusion process\par}
\smallskip

Multi-site symmetric simple exclusion processes extend the SSEP by imposing a finite-size constraint, whereby each particle prevents occupation of its $M$ nearest-neighboring sites. Their fluctuating hydrodynamic descriptions have the same form as that of the BHR, differing only through the transport coefficients~\cite{2004_Schonherr_Exclusion,2013_Krapivsky_Dynamics,2026_Saha_Bottom}, $D_{M}(\rho)=\tfrac{1}{(1-M\rho)^2}$ and $\sigma_{M}(\rho)=\frac{2\rho[1-(M+1)\rho]}{1-M\rho}$. Although the corresponding MSRJD path-integral \eqref{eq:path_integral_observable} leads to a difficult variational problem, the canonical transformation in \eqref{eq:transform_mft} extends directly to this case upon replacing $a$ by $M$. Under this transformation, the transport coefficients reduce to those of the SSEP, corresponding to the $M=0$ limit: $D_{0}(r)=1$ and $\sigma_{0}(r)=2r(1-r)$. Accordingly, the dynamical action \eqref{eq:mft_action}, initial free-energy functional \eqref{eq:initial_free_energy}, and relevant observables (\ref{eq:z_t_defn}-\ref{eq:Q_t_defn}) are mapped to their SSEP counterparts, thereby reducing the SCGF calculation for the multi-site exclusion process to the corresponding SSEP problem.

For the annealed ensemble, using known results for SSEP~\cite{2009_Derrida_Current,2017_Imamura_Large}, we obtain explicit expressions for the SCGF of the tracer-position and integrated-current. They take similar forms as in (\ref{eq:tracer_scgf}, \ref{eq:current_scgf}), with $a$ replaced by $M$, and with the function $\chi$ given by
\begin{equation}\label{eq:psi_multi_ssep_anneal}
\chi_{\mathcal{A}}(\xi,B)=\frac{\xi}{2}\!\ln\!\frac{1+\bar{r}_+(\mathrm{e}^{-B}-1)}{1+\bar{r}_-(\mathrm{e}^{B}-1)}-\sum_{n\ge1}\!\frac{(-\omega(B))^nI_n(\xi)}{n}
\end{equation}
where 
\begin{equation}
\omega(B)=\bar{r}_-(1-\bar{r}_+)(\mathrm{e}^{B}-1)+\bar{r}_+(1-\bar{r}_-)(\mathrm{e}^{-B}-1),
\end{equation}
with $I_n(\xi)=-\tfrac{\xi}{2}+\int_{-\xi}^{\infty}\mathrm{d}u\,H(\sqrt{n}u)$ and $\bar{r}_\mp=\tfrac{\bar{\rho}_\mp}{1-M\bar{\rho}_\mp}$. 

Fig.~\ref{fig:triple_ssep_anneal_current} presents a numerical verification of the annealed integrated-current SCGF, a result that had remained open~\cite{2023_Rizkallah_Duality}. In the quenched ensemble, however, due to unavailability of corresponding results in SSEP for arbitrary density, our explicit results are restricted to the uniform half-filled case~\cite{2009_Derrida_Current2,2015_Sadhu_Large}, $\bar{\rho}(x)=\tfrac{1}{2(M+1)}$, with the SCGFs satisfying $\mu_{\mathcal{A}}(\lambda)=\sqrt{2}\mu_{\mathcal{Q}}(\lambda)$.

\end{document}


\title{Supplement to `An exactly solvable macroscopic fluctuation theory of single-file diffusion'}

\author{Sandeep Jangid}
\affiliation{Department of Theoretical Physics, Tata Institute of Fundamental Research, Homi Bhabha Road, Mumbai 400005, India.}

\author{Soumyabrata Saha}
\affiliation{Department of Theoretical Physics, Tata Institute of Fundamental Research, Homi Bhabha Road, Mumbai 400005, India.}

\author{Kapil Sharma}
\affiliation{Department of Theoretical Physics, Tata Institute of Fundamental Research, Homi Bhabha Road, Mumbai 400005, India.}

\author{Jitendra Kethepalli}
\affiliation{Laboratoire de Physique Th\'eorique et Mod\'elisation, CNRS UMR 8089, CY Cergy Paris Universit\'e, 95302 Cergy-Pontoise Cedex, France.}
\affiliation{JEIP, UAR 3573 CNRS, Collège de France, PSL Research University, 11 Place Marcelin Berthelot, 75321 Paris Cedex 05, France.}

\author{Benjamin Guiselin}
\affiliation{Laboratoire Charles Coulomb (L2C), Universit\'e de Montpellier, CNRS, Montpellier, France.}

\author{Jacopo De Nardis}
\affiliation{Laboratoire de Physique Th\'eorique et Mod\'elisation, CNRS UMR 8089, CY Cergy Paris Universit\'e, 95302 Cergy-Pontoise Cedex, France.}
\affiliation{JEIP, UAR 3573 CNRS, Collège de France, PSL Research University, 11 Place Marcelin Berthelot, 75321 Paris Cedex 05, France.}

\author{Tridib Sadhu}
\affiliation{Department of Theoretical Physics, Tata Institute of Fundamental Research, Homi Bhabha Road, Mumbai 400005, India.}

\maketitle

\tableofcontents

\section{The point-particle mapping}
We use the free-volume coordinate centered on the tracer, defined as
\begin{equation}\label{eq:coordinate_transform}
X(x,s)=x-a\int_{z(s)}^{x}\mathrm{d}y\,\rho(y,s)
\end{equation}
The Jacobian of the free-volume coordinate transformation is given by the spatial derivative
\begin{equation}\label{eq:spatial_derivative}
\partial_xX=1-a\rho(x,s)
\end{equation}
which yields $\mathrm{d}X=(1-a\rho)\mathrm{d}x$. The conservation of number of particles in an interval $\rho(x,s)\mathrm{d}x=r(X,s)\mathrm{d}X$ then leads to
\begin{equation}\label{eq:density_transform}
r(X,s)=\frac{\rho(x,s)}{1-a\rho(x,s)}\implies\rho(x,s)=\frac{r(X,s)}{1+ar(X,s)}
\end{equation}
Consequently, the chain-rule of derivatives give the useful relation
\begin{equation}\label{eq:chain_rule}
\partial_x=\frac{\partial_X}{1+ar}
\end{equation}

The velocity of the free-volume coordinate follows from the temporal derivative
\begin{equation}\label{eq:temporal_derivative_1st}
\partial_sX=a\bigg[\dot{z}(s)\,\rho(z(s),s)-\int_{z(s)}^{x}\mathrm{d}x'\partial_s\rho(x',s)\bigg].
\end{equation}
Since the rank of the tracer is preserved at all times, we have $\tfrac{\mathrm{d}}{\mathrm{d}s}\int_{-\infty}^{z(s)}\mathrm{d}x'\rho(x',s)=0$ and using the continuity equation, we write
\begin{equation}
\dot{z}(s)\,\rho(z(s),s)=-\int_{-\infty}^{z(s)}\mathrm{d}x'\partial_s\rho(x',s)
\end{equation}
Using this in \eqref{eq:temporal_derivative_1st}, we arrive at
\begin{align}
\partial_sX&=-a\bigg[\int_{-\infty}^{z(s)}\mathrm{d}x'\partial_s\rho(x',s)+\int_{z(s)}^{x}\mathrm{d}x'\partial_s\rho(x',s)\bigg]\nonumber\\
&=a\int_{-\infty}^{x}\mathrm{d}x'\partial_{x'}J_{\rho}(x',s)
\end{align}
where we used the continuity equation, $\partial_s\rho=-\partial_xJ_{\rho}$ we write. Performing the integration and using $J_{\rho}(-\infty,s)=0$., we finally obtain
\begin{equation}\label{eq:temporal_derivative}
\partial_sX(x,s)=aJ_{\rho}(x,s).
\end{equation}

\subsection{Transformation of the fluctuating hydrodynamic equation}
The fluctuating hydrodynamic equation for the BHR describes the noisy evolution of the density $\rho(x,s)$ in terms of the two transport coefficients, $D_a(\rho)=\tfrac{1}{(1-a\rho)^2}$ and $\sigma_a(\rho)=2\rho$,
\begin{equation}\label{eq:noisy_hydro}
\partial_s\rho(x,s)=-\partial_xJ_\rho(x,s)\;,\;\;J_\rho(x,s)=-D_a(\rho(x,s))\partial_x\rho(x,s)+\frac{\sqrt{\sigma_a(\rho(x,s))}}{t^{1/4}}\eta_\rho(x,s)
\end{equation}
where, $\left<\eta_\rho(x,s)\eta_\rho(x',s')\right>=\delta(x-x')\delta(s-s')$.

Under the free-volume coordinate mapping with (\ref{eq:spatial_derivative}, \ref{eq:density_transform}), the noisy current transforms as
\begin{equation}
J_\rho(x,s)\equiv-(1+ar(X,s))^2\frac{1}{(1+ar(X,s))^3}\partial_Xr(X,s)+\frac{1}{t^{1/4}}\sqrt{\frac{2r(X,s)}{1+ar(X,s)}}\frac{1}{\sqrt{1+ar(X,s)}}\eta_r(X,s).
\end{equation}
For transforming the noise, we used the preservation of the normalization of the Dirac-delta function under the transformation, $\int\mathrm{d}x\,\delta(x-x')\equiv\int\mathrm{d}X(1+ar(X,s))\delta(x(X,s)-x(X',s))=\int\mathrm{d}X\delta(X-X')$. This leads to $\left<\eta_\rho(x,s)\eta_\rho(x',s')\right>=\tfrac{1}{1+ar(X,s)}\delta(X-X')\delta(s-s')$ and thus, the unit Gaussian-white noise in the free-volume coordinate is given by $\eta_\rho\equiv\tfrac{1}{\sqrt{1+ar}}\eta_r$. The above simplifies to give
\begin{align}
J_\rho(x,s)&\equiv\frac{1}{(1+ar(X,s))}\bigg(-\partial_Xr(X,s)+\frac{\sqrt{2r(X,s)}}{t^{1/4}}\eta_r(X,s)\bigg),\nonumber\\
\implies J_\rho(x,s)&\equiv\frac{J_r(X,s)}{1+ar(X,s)}\;,\;\;J_r(X,s)=-\partial_Xr(X,s)+\frac{\sqrt{2r(X,s)}}{t^{1/4}}\eta_r(X,s)\label{eq:current_reln}
\end{align}
Thus, the RHS of \eqref{eq:noisy_hydro} becomes
\begin{align}
-\partial_xJ_\rho(x,s)&\equiv-\frac{1}{1+ar(X,s)}\partial_X\bigg(\frac{J_r(X,s)}{1+ar(X,s)}\bigg)\nonumber\\
&=-\frac{1}{(1+ar(X,s))^2}\bigg(\partial_XJ_r(X,s)-\frac{aJ_r(X,s)\partial_Xr(X,s)}{1+ar(X,s)}\bigg)\label{eq:rhs_trans}
\end{align}

Now, under the free-volume coordinate transformation, the LHS of \eqref{eq:noisy_hydro} becomes
\begin{align}
\partial_s\rho&\equiv\frac{1}{(1+ar(X,s))^2}\big(\partial_sr(X,s)+\partial_sX(x,s)\partial_Xr(X,s)\big)\nonumber\\
&=\frac{1}{(1+ar(X,s))^2}\bigg(\partial_sr(X,s)+\frac{aJ_r(X,s)\partial_Xr(X,s)}{1+ar(X,s)}\bigg)\label{eq:lhs_trans}
\end{align}
where we used (\ref{eq:temporal_derivative}, \ref{eq:current_reln}).

Finally, combining \eqref{eq:lhs_trans} with \eqref{eq:rhs_trans}, we arrive at
\begin{equation}
\partial_sr(X,s)=-\partial_XJ_r(X,s)\;,\;\;J_r(X,s)=-\partial_Xr(X,s)+\frac{\sqrt{2r(X,s)}}{t^{1/4}}\eta_r(X,s)
\end{equation}
with $\left<\eta_r(X,s)\eta_r(X',s')\right>=\delta(X-X')\delta(s-s')$,which is the fluctuating hydrodynamic equation of point-particle.

\subsection{Transformation of observables}
\textit{Tracer-position}---The tracer-position, $z_t[\rho]$ is implicitly defined as
\begin{equation}\label{eq:tracer_defn}
\int_{0}^{z_t[\rho]}\mathrm{d}x\,\rho(x,t)=\int_{0}^{\infty}\mathrm{d}x\big(\rho(x,t)-\rho(x,0)\big)
\end{equation}
At $s=t$, from \eqref{eq:coordinate_transform} we have
\begin{equation}\label{eq:coord_i}
X(z_t[\rho],t)=z_t[\rho]\equiv\widetilde{z}_t[r]
\end{equation}
while $X(0,t)\equiv\mathcal{Z}=a\int_{0}^{z_t}\mathrm{d}x'\rho(x',t)$. This simplifies on using the definition of tracer-position \eqref{eq:tracer_defn} and noting that the RHS is the integrated-current (see \eqref{eq:current_defn}), to give
\begin{equation}\label{eq:Zcal_defn}
\mathcal{Z}=aQ_t[\rho]\equiv a\widetilde{Q}_t[r]
\end{equation}
Similarly at $s=0$, we have $X(0,0)=0$. Thus, \eqref{eq:tracer_defn} becomes
\begin{equation}\label{eq:tracer_transform}
\int_{0}^{\widetilde{z}_t[r]}\mathrm{d}X\,r(X,t)=\int_{0}^{\infty}\mathrm{d}X\big(r(X,t)-r(X,0)\big)
\end{equation}
where we added $\int_{0}^{\mathcal{Z}}\mathrm{d}X\,r(X,t)$ to both sides. This implies that the tracer-position $z_t[\rho]$ maps to the tracer-position in the point-particle system as defined implicitly in the above equatio.

\textit{Integrated-current}---The integrated-current, $Q_t[\rho]$ is defined as
\begin{equation}\label{eq:current_defn}
Q_t[\rho]=\int_{0}^{\infty}\mathrm{d}x\big(\rho(x,t)-\rho(x,0)\big)
\end{equation}
Following similar steps as in the tracer-position, we find that the above becomes
\begin{equation}\label{eq:current_transform}
\widetilde{Q}_t[r]=\int_{0}^{\infty}\mathrm{d}X\big(r(X,t)-r(X,0)\big)-\int_{0}^{\mathcal{Z}}\mathrm{d}X\,r(X,t)
\end{equation}
where, $\mathcal{Z}$ is determined implicitly from \eqref{eq:Zcal_defn}. Thus, the integrated-current $Q_t[\rho]$ maps to the observable \eqref{eq:current_transform} subject to the constraint \eqref{eq:Zcal_defn}.

\textit{Local-height}---The local-height observable measured at the position $\mathfrak{z}$ is defined as
\begin{equation}\label{eq:height_defn}
h_t^{(\mathfrak{z})}[\rho]=Q_t[\rho]-\int_{0}^{\mathfrak{z}}\mathrm{d}x\,\rho(x,t)
\end{equation}
which becomes under the point-particle mapping, after some simplification,
\begin{equation}\label{eq:height_transform}
\widetilde{h}_t^{\mathfrak{Z}}[\rho]=Q_t[r]-\int_{0}^{\mathfrak{Z}}\mathrm{d}X\,r(X,t)
\end{equation}
where $X(\mathfrak{z},t)\equiv\mathfrak{Z}=\mathfrak{z}-\int_{z_t}^{\mathfrak{z}}\mathrm{d}x\,\rho(x,t)$. Combining the definitions of the integrated-current with that of the tracer-position, we find
\begin{equation}\label{eq:Zfrak_defn}
\mathfrak{Z}-\mathfrak{z}=ah_t^{\mathfrak{z}}[\rho]\equiv a\widetilde{h}_t^{\mathfrak{Z}}[r]
\end{equation}
Thus, the local-height $h_t^{(\mathfrak{z})}[\rho]$ maps to the local-height measured at position $\mathfrak{Z}$ subject to the condition \eqref{eq:Zfrak_defn}.

Note that the transformed integrated-current \eqref{eq:current_transform} is the local-height measured at position $\mathcal{Z}$ (i.e., $\widetilde{Q}_t[r]\equiv\widetilde{h}_t^{(\mathcal{Z})}[r]$) with $\mathcal{Z}$ self-consistently following from the observable itself.

\section{The relation between local-height and tracer-position}

The tracer-position (implicitly defined in \eqref{eq:tracer_defn}) follows from the local-height implicitly
\begin{equation}\label{eq:tracer_height_reln}
z_t[\rho]=\mathfrak{Z}\Leftrightarrow h_t^{(\mathfrak{Z})}[\rho]=0
\end{equation}
The LDF governing the local-height fluctuations is defined as
\begin{equation}
\mathcal{P}\bigg(\frac{h_t^{(\mathfrak{Z})}}{\sqrt{t}}=\kappa\bigg)\asymp\mathrm{e}^{-\sqrt{t}\phi^{\rm lh}(\zeta,\kappa)}
\end{equation}
where $\tfrac{\mathfrak{Z}}{\sqrt{t}}=\zeta$, while the corresponding SCGF is
\begin{equation}
\big<\mathrm{e}^{\lambda h_t^{(\mathfrak{Z})}}\big>\asymp\mathrm{e}^{\sqrt{t}\mu^{\rm lh}(\zeta,\lambda)}.
\end{equation}
These are related through a Legendre-transform
\begin{equation}
\phi^{\rm lh}(\zeta,\kappa)=\max_{\lambda}\big(\lambda\kappa-\mu^{\rm lh}(\zeta,\lambda)\big)
\end{equation}

From \eqref{eq:tracer_height_reln}, then the LDF governing tracer-position fluctuations is given by
\begin{equation}\label{eq:tracer_ldf_height_scgf}
\phi^{\rm tp}(\xi)=-\mu^{\rm lh}(\xi,\lambda)\;,\;\;\partial_\lambda\mu^{\rm lh}(\xi,\lambda)=0.
\end{equation}
The SCGF for the tracer-position is then obtained as $\mu^{\rm tp}(\lambda)=\lambda\xi+\mu^{\rm lh}(\xi,B)$, with $\partial_B\mu^{\rm lh}(\xi,B)=0$ and $\partial_\xi\mu^{\rm lh}(\xi,B)=-\lambda$.

Note that the integrated-current is straightforwardly obtained from the local-height as
\begin{equation}\label{eq:current_height_reln}
Q_t[\rho]=h_t^{(0)}[\rho]
\end{equation}
such that the SCGF for the integrated-current is given by
\begin{equation}\label{eq:current_scgf_height_scgf}
\mu^{\rm ic}(\lambda)=\mu^{\rm lh}(0,\lambda).
\end{equation}
and the corresponding LDF takes the form, $\phi^{\rm ic}(j)=\lambda j-\mu^{\rm lh}(0,\lambda)$ with $\partial_\lambda\mu^{\rm lh}(0,\lambda)=j$.

\subsection{Local-height statistics in the BHR}
The generating-function of the local-height measured at $\mathfrak{z}$ in the BHR can be written as a path-integral
\begin{equation}
\left<\mathrm{e}^{\lambda h_t^{\mathfrak{z}}[\rho]}\right>=\int\mathrm{d}\mathfrak{Z}\,\mathrm{e}^{\lambda\widetilde{h}_t^{(\mathfrak{Z})}[r]}\delta\big(\mathfrak{Z}-\mathfrak{z}-a\widetilde{h}_t^{(\mathfrak{Z})}[r]\big)
\end{equation}
where the self-consistency condition on $\mathfrak{Z}$ in \eqref{eq:Zfrak_defn} enforced through a delta-functional. Exponentiating the delta-functional by introducing an auxiliary parameter $B$, the above becomes
\begin{equation}
\Big<\mathrm{e}^{\lambda h_t^{\mathfrak{Z}}[\rho]}\Big>=\int\mathrm{d}\mathcal{Z}\,\mathrm{d}B\,\mathrm{e}^{(\mathfrak{Z}-\mathfrak{z})B}\Big<\mathrm{e}^{(\lambda-aB)\widetilde{h}_t^{(\mathfrak{Z})}[r]}\Big>
\end{equation}
In the large-$t$ limit, we can perform a saddle-point calculation of the path-integral. The average in angular brackets is the generating-function of the local-height observable measured at position $\mathfrak{Z}$ in the point-particle system, evaluated at fugacity $\lambda-aB$. Denoting the rescaled positions $\tfrac{\mathfrak{z}}{\sqrt{t}}=\xi$ and $\tfrac{\mathfrak{Z}}{\sqrt{t}}=\Xi$, we arrive at
\begin{align}
\mu^{\rm lh}(\xi,\lambda)=&\,(\Xi-\xi)B+\chi(\Xi,\lambda-aB),\label{eq:height_scgf_bhr}\\
&\partial_\Xi\chi(\Xi,\lambda-aB)=-B,\\
&\partial_B\chi(\Xi,\lambda-aB)=\xi-\Xi
\end{align}
where $\chi(\Xi,\lambda-aB)$ denotes the local-height SCGF for point-particles.

Note that the computation of the integrated-current statistics in the BHR follows similarly where the delta-functional now enforces the constraint \eqref{eq:Zcal_defn}. This is because of the previously discussed realization that in the point-particle representation, it transforms to a local-height observable measured at $\tfrac{\mathcal{Z}}{\sqrt{t}}=\xi$.

In the annealed ensemble, the local-height SCGF of the BHR satisfies a fluctuation symmetry relation
\begin{align}
&\mu_\mathcal{A}^{\rm lh}(\xi,\lambda)=\mu_\mathcal{A}^{\rm lh}(-\xi,c_0-\lambda)-c_1\xi,\\
&c_n=\int_{\rho_-}^{\rho_+}\mathrm{d}w\,w^n\frac{2D_a(w)}{\sigma_a(w)}
\end{align}
Using the tracer-height correspondence \eqref{eq:tracer_ldf_height_scgf}, and noting that $\lambda\equiv c_0-\lambda$ is a one-to-one reparametrization of the extremized variable, we arrive at
\begin{equation}
\phi_\mathcal{A}^{\rm tp}(\xi)-\phi_\mathcal{A}^{\rm tp}(-\xi)=c_1\xi.
\end{equation}
Similarly, from the current-height relation \eqref{eq:current_scgf_height_scgf}, we obtain
\begin{equation}
\mu_\mathcal{A}^{\rm ic}(\lambda)=\mu_\mathcal{A}^{\rm ic}(c_0-\lambda)
\end{equation}

\subsection{Tracer-position and integrated-current statistics from the local-height SCGF}
Using \eqref{eq:height_scgf_bhr} in \eqref{eq:tracer_ldf_height_scgf}, we obtain the tracer-position LDF for the BHR as
\begin{equation}
\phi^{\rm tp}(\xi)=-(\Xi-\xi)B-\chi(\Xi,\Lambda-aB)
\end{equation}
where the RHS is extremized over $\Xi$, $B$ and $\Lambda$. The corresponding SCGF is then given by
\begin{equation}
\mu^{\rm tp}(\lambda)=\operatorname*{extr}_{\xi,\Xi,B,\Lambda}\big[\Xi B+(\lambda-B)\xi+\chi(\Xi,\Lambda-aB)\big].
\end{equation}
The saddle over $\xi$ imposes $B=\lambda$, giving $\mu^{\rm tp}(\lambda)=\operatorname*{extr}_{\Xi,\Lambda}\left(\Xi\lambda+\chi(\Xi,\Lambda-a\lambda)\right)$. Since, $\Lambda$ is itself an extremized variable, we define $\beta=\Lambda-a\lambda$, and write
\begin{equation}
\mu^{\rm tp}(\lambda)=\operatorname*{extr}_{\Xi,\beta}\big(\Xi\lambda+\chi(\Xi,\beta)\big)
\end{equation}
where $\Xi$ and $\beta$ are extremized. This is the announced result in the \emph{Letter}, with the relabeling $\Xi\to\xi$ and $\beta\to B$.

For the integrated-current, the calculation becomes much more straightforward. From \eqref{eq:current_height_reln}, we arrive at
\begin{equation}
\mu^{\rm ic}(\lambda)=\operatorname*{extr}_{\Xi,B}\big(\Xi B+\chi(\Xi,\lambda-aB)\big)
\end{equation}
which is the result reported in the \emph{Letter}, with the relabeling $\Xi\to\xi$.

\section{Solving the MFT equations of the BHR}
The MFT equations govern the optimal evolution of the hydrodynamic fields which are obtained from the saddle-point computation of the generating function of an observable
\begin{equation}
\big<\mathrm{e}^{\lambda\mathcal{O}_t}\big>=\int\mathcal{D}[\rho,\hat{\rho}]\mathrm{e}^{\lambda\mathcal{O}_t[\rho]-\sqrt{t}\left(\mathcal{F}_a[\rho(x,0)]+\mathcal{S}_a[\rho,\hat{\rho}]\right)}.
\end{equation}
For the BHR, we have
\begin{equation}\label{eq:ini_f_bhr}
\mathcal{F}_a[\rho(x,0)]=\int_{-\infty}^{\infty}\mathrm{d}x\int_{\bar{\rho}(x)}^{\rho(x,0)}\mathrm{d}w\,\frac{\rho(x,0)-w}{w(1-aw)^2},
\end{equation}
and
\begin{equation}\label{eq:dyn_act_bhr}
\mathcal{S}_a[\rho,\hat{\rho}]=\int_{0}^{1}\mathrm{d}s\int_{-\infty}^{\infty}\mathrm{d}x\bigg\{\hat{\rho}\,\partial_s\rho-\bigg[\rho\,\partial_x\hat{\rho}-\frac{\partial_x\rho}{(1-a\rho)^2}\bigg]\partial_x\hat{\rho}\bigg\}
\end{equation}
For the BHR in the annealed ensemble, these take the form
\begin{align}
\partial_s\rho&=\partial_x\bigg[\frac{\partial_x\rho}{(1-a\rho)^2}\bigg]-2\partial_x(\rho\partial_x\hat{\rho}),\label{eq:rho_mft}\\
\partial_s\hat{\rho}&=-\frac{\partial_x^2\hat{\rho}}{(1-a\rho)^2}-(\partial_x\hat{\rho})^2,\label{eq:rhohat_mft}
\end{align}
subject to the initial boundary condition
\begin{equation}\label{eq:initial_bc_anneal}
\hat{\rho}(x,0)=-\lambda\frac{\delta\mathcal{O}_t[\rho]}{\delta\rho(x,0)}+\int_{\bar{\rho}(x)}^{\rho(x,0)}\frac{\mathrm{d}w}{w(1-aw)^2},
\end{equation}
and the final boundary condition
\begin{equation}\label{eq:final_bc}
\hat{\rho}(x,1)=\lambda\frac{\delta\mathcal{O}_t[\rho]}{\delta\rho(x,1)}.
\end{equation}

\subsection{A canonical transformation of the hydrodynamic fields}
Solving these MFT equations is undoubtedly difficult. In order to make the problem simpler, we map them to the point-particle representation. The coordinate map \eqref{eq:coordinate_transform} and the corresponding density field map \eqref{eq:density_transform} has already been discussed. Accordingly, we make the canonical response field transformation
\begin{equation}\label{eq:response_transform}
\hat{r}(X,s)=\hat{\rho}(x,s)-a\int_{z(s)}^{x}\mathrm{d}x'\rho(x',s)\partial_{x'}\hat{\rho}(x',s)\implies\hat{\rho}(x,s)=\hat{r}(X,s)+a\int_{z(s)}^{X}\mathrm{d}X'r(X',s)\partial_{X'}\hat{r}(X',s)
\end{equation}
which preserves the gradient structure $\partial_x\hat{\rho}(x,s)=\partial_x\hat{r}(X,s)$. 

Consider the Hamiltonian part of the action. Using the Jacobian \eqref{eq:spatial_derivative} and the chain-rule \eqref{eq:chain_rule}, we have
\begin{align}
\mathcal{H}_a[\rho,\hat{\rho}]&\equiv\int_{-\infty}^{\infty}\mathrm{d}X(1+ar)\bigg[\frac{r}{1+ar}\,\partial_X\hat{r}-(1+ar)^2\frac{\partial_Xr}{(1+ar)^3}\bigg]\partial_X\hat{r}\nonumber\\
&=\int_{-\infty}^{\infty}\mathrm{d}X\big(r\,\partial_X\hat{r}-\partial_Xr\big)\partial_X\hat{r}\label{eq:hamilton_transform}
\end{align}

Next, we look at the kinetic-term. We take the time-derivative of density and write
\begin{align}
\partial_s\rho&\equiv\frac{1}{(1+ar)^2}(\partial_sr+\partial_sX\partial_Xr)\nonumber\\
&=\frac{\partial_sr}{(1+ar)^2}+\frac{a\mathcal{J}_r\,\partial_Xr}{(1+ar)^3}
\end{align}
where $\partial_sr=-\partial_X\mathcal{J}_r$. Using this, we get the kinetic-term
\begin{align}
\int_{-\infty}^{\infty}\mathrm{d}x\,\hat{\rho}\,\partial_s\rho&\equiv\int_{-\infty}^{\infty}\mathrm{d}X(1+ar)\,\hat{\rho}\bigg[\frac{\partial_sr}{(1+ar)^2}+\frac{a\mathcal{J}_r\,\partial_Xr}{(1+ar)^3}\bigg]\nonumber\\
&=\int_{-\infty}^{\infty}\mathrm{d}X\,\hat{r}\,\partial_sr+\int_{-\infty}^{\infty}\mathrm{d}X\,\bigg[\bigg(\frac{\hat{\rho}}{1+ar}-\hat{r}\bigg)\partial_sr+\frac{a\hat{\rho}\,\mathcal{J}_r\,\partial_Xr}{(1+ar)^2}\bigg]
\end{align}
Since, $\partial_X\!\left(\tfrac{\hat{\rho}}{1+ar}-\hat{r}\right)=-\tfrac{a\hat{\rho}\,\partial_Xr}{(1+ar)^2}$ and $\partial_sr=-\partial_X\mathcal{J}_r$, the integrand in the second integral can be written as a total derivative. This reduces the second integral to the spatial boundary term
\begin{equation}
\bigg[\mathcal{J}_r\bigg(\hat{r}-\frac{\hat{\rho}}{1+ar}\bigg)\bigg]_{-\infty}^{\infty}
\end{equation}
which vanishes since the fields approach time-independent constants as $X\to\pm\infty$. Finally, we obtain
\begin{equation}\label{eq:kinetic_transform}
\int_{-\infty}^{\infty}\mathrm{d}x\,\hat{\rho}\,\partial_s\rho\equiv\int_{-\infty}^{\infty}\mathrm{d}X\,\hat{r}\,\partial_sr
\end{equation}

Combining the Hamiltonian \eqref{eq:hamilton_transform} along with the above for the kinetic component, we get the transformed dynamical action \eqref{eq:dyn_act_bhr} in the free-volume coordinate as
\begin{equation}
S_0[r,\hat{r}]=\int_{0}^{1}\mathrm{d}s\int_{-\infty}^{\infty}\mathrm{d}X\big[\hat{r}\,\partial_sr-\big(r\,\partial_X\hat{r}-\partial_Xr\big)\partial_X\hat{r}\big]
\end{equation}

The free-energy functional \eqref{eq:ini_f_bhr} describing initial-density fluctuations transforms straightforwardly to the corresponding point-particle representation
\begin{equation}
\mathcal{F}_0[r(X,0)]=\int_{-\infty}^{\infty}\mathrm{d}X\int_{\bar{r}(X)}^{r(X,0)}\mathrm{d}w\,\frac{r(X,0)-w}{w}
\end{equation}
where we used $\mathrm{d}x=\mathrm{d}X(1+ar)$ and $\rho=\tfrac{r}{1+ar}$. Note that $\bar{r}(X)=\bar{r}_-\theta(-X)+\bar{r}_+\theta(X)$ with $\bar{r}_\mp=\tfrac{\bar{\rho}_\mp}{1-a\bar{\rho}_\mp}$, denotes the initial step-like point-particle density profile.

Thus, we finally arrive at the generating-function in the point-particle representation
\begin{equation}
\big<\mathrm{e}^{\lambda\widetilde{\mathcal{O}}_t}\big>=\int\mathcal{D}[r,\hat{r}]\mathrm{e}^{\lambda\widetilde{\mathcal{O}}_t[r]-\sqrt{t}\left(\mathcal{F}_0[r(X,0)]+\mathcal{S}_0[r,\hat{r}]\right)}
\end{equation}
which now corresponds to solving the MFT equations
\begin{align}
\partial_sr&=\partial_X^2r-2\partial_X(r\partial_X\hat{r}),\label{eq:r_mft}\\
\partial_s\hat{r}&=-\partial_X^2\hat{r}-(\partial_X\hat{r})^2,\label{eq:rhat_mft}\\
\hat{r}(X,0)&=-\lambda\frac{\delta\widetilde{\mathcal{O}}_t[r]}{\delta r(X,0)}+\int_{\bar{r}(X)}^{r(X,0)}\frac{\mathrm{d}w}{w},\label{eq:rhat_ini}\\
\hat{r}(X,1)&=\lambda\frac{\delta\widetilde{\mathcal{O}}_t[r]}{\delta r(X,1)}.\label{eq:rhat_fin}
\end{align}

Performing a canonical Cole-Hopf transformation $(r,\hat{r})\to(Q,P)=(r\mathrm{e}^{-\hat{r}},\mathrm{e}^{\hat{r}})$, (\ref{eq:r_mft}, \ref{eq:rhat_mft}) reduces to $\partial_sQ=\partial_X^2Q$ and $\partial_sP=-\partial_X^2P$ whose general solution readily follow: $Q(X,s)=\int_{-\infty}^{\infty}\mathrm{d}X'\mathcal{G}_s(X-X')Q(X,0)$ and $P(X,s)=\int_{-\infty}^{\infty}\mathrm{d}X'\mathcal{G}_{1-s}(X-X')P(X,1)$. Using these, we write a general solution for $(r,\hat{r})$ as
\begin{align}
r(X,s)&=\bigg[\int_{-\infty}^{\infty}\mathrm{d}X'\mathcal{G}_s(X-X')r(X',0)\mathrm{e}^{-\hat{r}(X',0)}\bigg]\bigg[\int_{-\infty}^{\infty}\mathrm{d}X'\mathcal{G}_{1-s}(X-X')\mathrm{e}^{\hat{r}(X',1)}\bigg]\\
\hat{r}(X,s)&=\ln\bigg[\int_{-\infty}^{\infty}\mathrm{d}X'\mathcal{G}_{1-s}(X-X')\mathrm{e}^{\hat{r}(X',1)}\bigg]
\end{align}
The above simplifies upon using (\ref{eq:rhat_ini}, \ref{eq:rhat_fin}) to yield
\begin{align}
r_{\mathcal{A}}(X,s)&=\mathrm{e}^{\hat{r}(X,s)}\int_{-\infty}^{\infty}\mathrm{d}X'\mathcal{G}_s(X-X')\bar{r}(X')\mathrm{e}^{\lambda\mathcal{V}_0(X')}\\
r_{\mathcal{Q}}(X,s)&=\mathrm{e}^{\hat{r}(X,s)}\int_{-\infty}^{\infty}\mathrm{d}X'\frac{\mathcal{G}_s(X-X')\bar{r}(X')}{\int_{-\infty}^{\infty}\mathrm{d}X''\mathcal{G}_{1}(X'-X'')\mathrm{e}^{\lambda\mathcal{V}_1(X'')}}\\
\hat{r}(X,s)&=\ln\int_{-\infty}^{\infty}\mathrm{d}X'\mathcal{G}_{1-s}(X-X')\mathrm{e}^{\lambda\mathcal{V}_1(X')}
\end{align}
where we denoted $\mathcal{V}_s(X')=\frac{\delta\widetilde{\mathcal{O}}_t[r]}{\delta r(X',s)}$, for $s=0,1$. They take the form
\begin{numcases}
{\mathcal{V}_0(X')=}
-\frac{\theta(X')}{r(\xi,1)}&for tracer-position\\
-\frac{\theta(X')}{1+ar(aj,1)}&for integrated-current\\
-\frac{\theta(X')}{1+ar(\zeta+a\kappa,1)}&for local-height
\end{numcases}
and
\begin{numcases}
{\mathcal{V}_1(X')=}
\frac{\theta(X'-\xi)}{r(\xi,1)}&for tracer-position\\
\frac{\theta(X'-aj)}{1+ar(aj,1)}&for integrated-current\\
\frac{\theta(X'-\zeta-a\kappa)}{1+ar(\zeta+a\kappa,1)}&for local-height
\end{numcases}

\section{Cumulants for non-equilibrium initial conditions}
The mean value of the tracer-position is
\begin{equation}
\frac{\big<z_t\big>}{\sqrt{t}}=\xi_0,
\end{equation}
where $\xi_0$ is the solution of
\begin{equation}
\xi_0=\frac{\bar{r}_--\bar{r}_+}{\bar{r}_+}\int_{\xi_0}^{\infty}\mathrm{d}u\,H(u).
\end{equation}

Similarly, the mean value of the integrated-current is
\begin{equation}
\frac{\big<Q_t\big>}{\sqrt{t}}=j_0,
\end{equation}
where $j_0$ is the solution of
\begin{equation}
j_0=\frac{\bar{r}_--\bar{r}_+}{1+a\bar{r}_+}\int_{aj_0}^{\infty}\mathrm{d}u\,H(u).
\end{equation}

The variances of tracer-position and integrated-current are respectively given by
\begin{align}
\frac{\big<z_t^2\big>}{\sqrt{t}}&=\frac{K(\xi_0)}{[L(\xi_0)]^2},\\
\frac{\big<Q_t^2\big>}{\sqrt{t}}&=\frac{K(aj_0)}{[1+aL(aj_0)]^2}
\end{align}
with
\begin{align}
K_\mathcal{A}(x)&=\bar{r}_-\int_{x}^{\infty}\mathrm{d}u\,H(u)+\bar{r}_+\int_{-x}^{\infty}\mathrm{d}u\,H(u)\\
K_\mathcal{Q}(x)&=\bar{r}_-\int_{x}^{\infty}\mathrm{d}u\,H(u)H(-u)+\bar{r}_+\int_{-x}^{\infty}\mathrm{d}u\,H(u)H(-u)\\
L(x)&=\bar{r}_-H(x)+\bar{r}_+H(-x)
\end{align}